\documentclass[twocolumn, prl]{revtex4}
\pdfoutput=1
\usepackage{graphicx}
\usepackage{epsfig}
\begin{document}
\title{Interaction-induced excited-band condensate in a double-well optical lattice }
\author{Qi Zhou,$^{1,2}$  J. V. Porto,$^{1}$ and S. Das Sarma$^{1,2}$}
\affiliation{$^{1}$Joint Quantum Institute and $^{2}$Condensed Matter Theory Center, Department
of Physics, University of Maryland, College Park, MD 20742}
\date{\today}
\begin{abstract}
We show theoretically that interaction effects in a double-well optical lattice can induce condensates in an excited band. For a symmetric double-well lattice, bosons condense into the bottom of the excited band at the edge of the Brillouin Zone if the chemical potential is above a critical value. For an asymmetric lattice, a condensate with zero momentum is automatically induced in the excited band by the condensate in the lowest band.   This is due to a combined effect of interaction and lattice potential, which reduces the band gap and breaks the inversion symmetry. Our work can be generalized to a superlattice composed of multiple-well potentials at each lattice site, where condensates can be induced in even higher bands.

 \end{abstract}
\maketitle\textit{}

Interaction between different orbitals lies at the heart of many challenging condensed matter phenomena such as heavy fermions and multi-band superconductors. Inspired by experimental developments in cold atom physics in the lowest band of optical lattices, physicists have begun to explore inter-band phenomena in optical lattices\cite{trey, Bloch1, interband1, interband2, Ernst, QZ}. It is hoped that the precise controllability of optical lattices can provide unique opportunities to reveal the interplay between multi-band physics and interaction effects in strongly correlated systems.  Although the specific results obtained in the multi-band optical lattice phenomena in these papers pertain to bosons, the generic considerations may also apply to fermions. 

 An important question regarding the multi-band physics in optical lattices is how to populate the atoms in excited bands. Bosons have been excited to the $p$-band of a square lattice, where they establish coherence along specific directions\cite{Bloch}. Very recently, a novel technique was used to populate bosons in the excited bands in a bipartite optical lattice\cite{Hem}. $p$- and $f$- band condensates were produced with life time up to a few tens of $ms$. These exciting experimental developments\cite{Bloch, Hem} should enable the study of several interesting phenomena  in higher bands of optical lattices recently predicted in the literature\cite{Sankar, CJW1, CJW2, Girvin, HZ}.  

We emphasize that the condensates in excited bands obtained in current experiments are not the ground state of the system, and they eventually decay. It is then desirable to have a scheme for producing condensates in excited bands that have much longer life time.  In this paper, we propose such a scheme using optical double-well lattices to produce excited-band condensates on top of an ordinary lowest-band condensate. The main results of this work are summarized as following. 

(I). In a symmetric double-well lattice, a condensate forms at the edge of the Brillouin Zone(BZ) in the lowest excited band with a finite momentum ${\bf k}$, if  the chemical potential is above a critical value that is achievable when the intra-band interaction is larger than the inter-band interaction. 

(II). In an asymmetric double-well lattice where the inversion symmetery is broken,  a stable condensate automatically forms at the top of the excited band with ${\bf k}=0$ in the presence of a lowest band condensate.  Unlike the case in (I), there is no threshold for this condensate to emerge and it can be induced by any finite condensate in the lowest band.

(III). Both the condensates discussed in (I) and (II) are the ground state properties of the system. They are stable and do not suffer a short life time problem.


(IV). Our studies on the double-well lattice can be generalized to other multiple-well (more than two) superlattices, for which even higher band condensates can be produced. 


{\bf (A) System and Hamiltonian:} A double-well lattice can be produced by adding a short lattice of twice the periodicity along the $x$ direction to the standard cubic lattice\cite{Bloch2}. Physics discussed here can be easily generalized to other lattice geometries\cite{La}. The general form of the potential for the case we consider here can be written as 
\begin{equation}
V({\bf R})=\left(\sum_{i}V_{L}^i\sin^2(\frac{\pi R_i}{d})\right)-V_S\sin^2(\frac{2\pi R_x}{d}+\frac{\varphi}{2}),\label{lp}
\end{equation}
where $i=x,y,z$, $d$ is the lattice spacing, $V_L^{i}$ is the amplitude of the long lattice along different directions and $V_S$ is the the amplitude of the short lattice along the $x$ direction. $\varphi$ characterizes the relative position between the two lattices. The symmetric case corresponds to $\varphi=0$ or $2\pi$. 

When $V_S$ is large enough, the lowest two bands are well separated from all the other ones. The many-body Hamiltonian in momentum space  can be written as 

\begin{equation}
H=\sum_{\sigma {\bf k}}\epsilon_{\sigma{\bf k}}\hat{a}^\dagger_{\sigma{\bf k}}\hat{a}_{\sigma{\bf k}}+\sum_{\sigma_i{\bf k}_i } \mathcal{U}^{\sigma_1\sigma_2\sigma_3\sigma_4}_{{\bf k_1}{\bf k_2}{\bf k_3}{\bf k_4}} \hat{a}^\dagger_{{\sigma_1}{\bf k}_1}\hat{a}^\dagger_{{\sigma_2}{\bf k}_2}\hat{a}_{{\sigma_3}{\bf k}_3}\hat{a}_{{\sigma_4}{\bf k}_4},
\end{equation}
where $\sigma=g,e$ are the indices for the lowest two bands,  $\epsilon_{\sigma{\bf k}}=\xi_{\sigma{\bf k}}-\mu$, $\xi_{\sigma{\bf k}}$ is the single particle spectrum of each band and ${\bf k}$ is defined in the first BZ.  $\mathcal{U}^{\sigma_1\sigma_2\sigma_3\sigma_4}_{{\bf k_1}{\bf k_2}{\bf k_3}{\bf k_4}}=\frac{2\pi \hbar^2a_s}{M}\int d{\bf R}\Psi^*_{\sigma_1{\bf k}_1}({\bf R})\Psi^*_{\sigma_2{\bf k}_2}({\bf R})\Psi_{\sigma_3{\bf k}_3}({\bf R})\Psi_{\sigma_4{\bf k}_4}({\bf R})\delta({\bf k}_1+{\bf k}_2-{\bf k}_3-{\bf k}_4={\bf K})$, ${\bf K}=2\pi/d(l_1,l_2,l_3)$ is the reciprocal lattice vector, $\Psi_{\sigma{\bf k}}({\bf R})=e^{i{\bf k}\cdot{\bf R}}u_{\sigma {\bf k}}({\bf R})$ is the Bloch wave function with crystal momentum ${\bf k}$ at band $\sigma$, $a_s$ is the scattering length and $M$ is the mass of the bosons. 

\begin{figure}[tbp]
\begin{center}
\includegraphics[width=2.4 in]{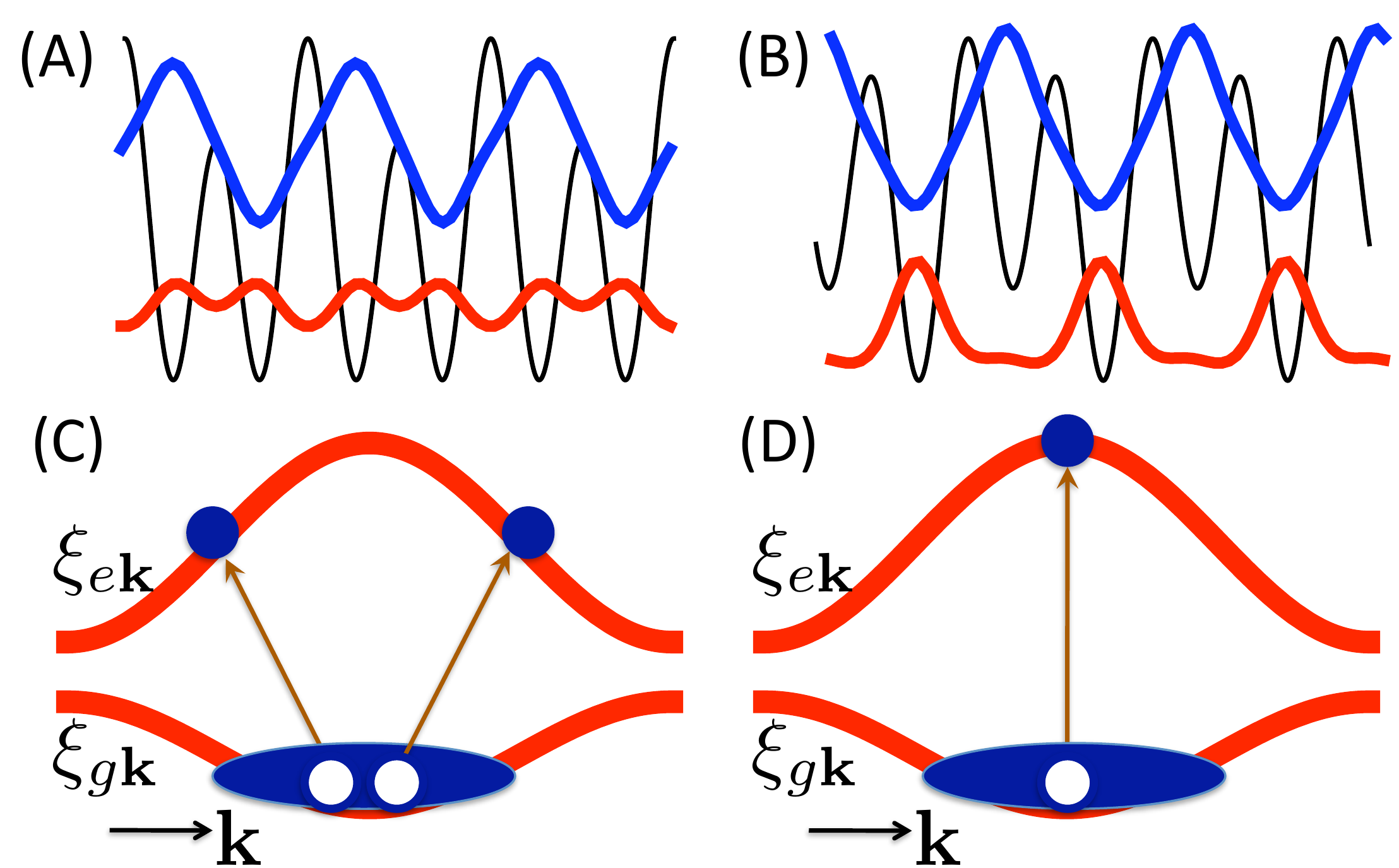}
\end{center}
\caption{(Color online) (A) and (B) are schematic of a symmetric and an asymmetric double-well lattice respectively. The Bloch wave functions of the lowest two bands at ${\bf k}=0$ are shown in red(bottom) and blue(top). (C) and (D) represent the pairing scattering $W_{\bf k}^*=\mathcal{U}^{eegg}_{{\bf k -k}00} $and the interaction induced inter-band hybridization $G_0^{'*}=2\mathcal{U}^{eggg}_{0000}$, which creates two particles at $({\bf k}, -{\bf k})$ and one particle at ${\bf k}=0$ in the excited band respectively. The blue cloud represents the lowest-band condensate, and the holes represent the particles annihilated. }\label{fig: fig1}
\end{figure}

To be concrete, we focus on the Gross-Pitaevski regime, i.e., we assume there is a condensate at the bottom of the lowest band $\langle \hat{a}_{g0}\rangle=\psi_g$. Because of the finite value of the gap,  the higher band condensate is usually much weaker than the one in the lowest band. To the zeroth order, the magnitude of $\psi_g$ can be determined by $\psi_g=\left(\frac{\mu-\xi_{g0}}{2U_g}\right)^{1/2}$,  where  $U_g=\mathcal{U}^{gggg}_{0000}$. For the excited band, there are many terms involved. Up to the order of $\psi_g^3$ and $\psi_g^2$, the Hamiltonian for the excited band can be written as $H_e=\sum_{{\bf k}}\tilde{\epsilon}_{e{\bf k}}\hat{a}^\dagger_{e{\bf k}}\hat{a}_{e{\bf k}}+U+U'+\sum_{\bf k\neq 0}({W}'_{\bf k}(\hat{a}_{e{\bf k}}\hat{a}_{g{-\bf k}}+\hat{a}_{g{\bf k}}\hat{a}_{e{-\bf k}})+c.c)$ , 
\begin{equation}
U=\psi_g^2\sum_{{\bf k}}\left({W}_{\bf k}\hat{a}_{e{\bf k}}\hat{a}_{e{-\bf k}}+c.c\right),\label{He}
\end{equation}
where $\tilde{\epsilon}_{e{\bf k}}=\epsilon_{e{\bf k}}+4\psi_g^2\mathcal{U}^{ggee}_{00{\bf k k}}$, $W_{\bf k}=\mathcal{U}^{ggee}_{00{\bf k -k}}$, ${W}'_{\bf k}=U^{ggge}_{00{\bf k}-{\bf k}}$ and 
\begin{equation}
U'=\psi_g^3({G}_0^*\hat{a}_{e0}^\dagger+{G}_0\hat{a}_{e0}),\label{Ge}
\end{equation}
where  $G_0=2\mathcal{U}^{ggge}_{0000}$. Because of momentum conservation, $U'$ only contains $\hat{a}_{e0}^\dagger$ or $\hat{a}_{e0}$. These leading order terms are shown in Fig.(\ref{fig: fig1}). The effect of higher order terms for determining the exact value of the excited-band condensate amplitude will be discussed in section (B) and (C). 

\begin{figure}[tbp]
\begin{center}
\includegraphics[width=3.2 in]{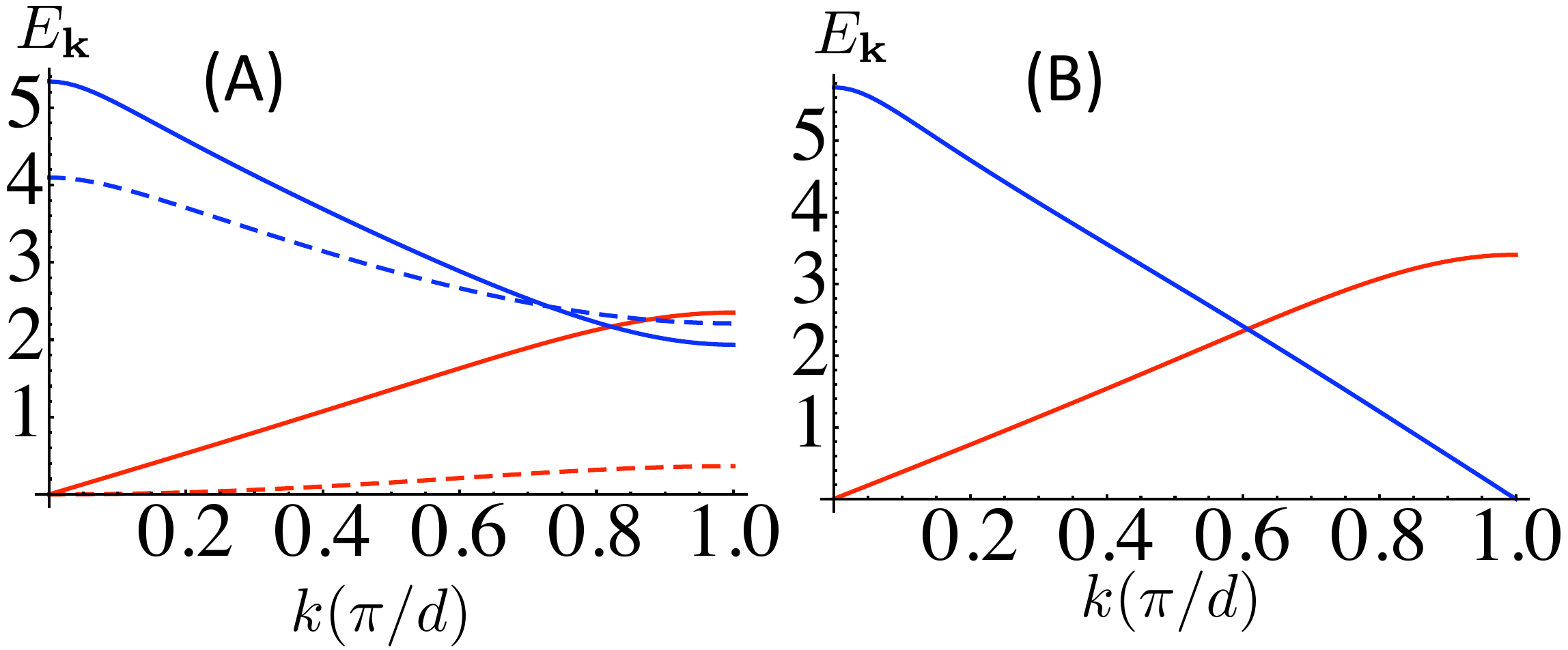}
\end{center}
\caption{(Color online) The excitation spectrum for the ground(bottom from left) and the first excited band(top from left) of a symmetric double-well lattice for (A) $\mu<\mu_c$ and (B) $\mu=\mu_c$, where $V_L^y=V_L^z=15E_R$, $V_L^x=4E_R$ $V_S=1E_R$ and $a_s/d=0.06$. $\psi_g^2=5$ at $\mu_c$ . Dashed lines in (A) represent the spectrum for non-interacting case. The ground band is also treated by Bogoliubov theory.}\label{fig: fig2}
\end{figure}

{\bf (B) Symmetric lattices:} Due to different parity of the wave function $u_{g0}(R)$ and $u_{e0}(R)$ in a symmetric double-well lattice, $G_0$ vanishes, similar to the discussion in reference\cite{QZ}. We also find numerically that $W'_{\bf k}$ is generally one or two orders of magnitude smaller than $W_{\bf k}$ for a broad parameter regions of lattice depth, i.e., $V_L$, $V_S$ $\sim 2-10 E_R$. Therefore we ignore $W'_{\bf k}$ in the following discussions.  Under this situation, $H_e$ is exactly the same as in the usual Bogoliubov theory for the excited band, i.e., $H_e=\sum_{{\bf k}}\tilde{\epsilon}_{e{\bf k}}\hat{a}^\dagger_{e{\bf k}}\hat{a}_{e{\bf k}}+\psi_g^2\sum_{{\bf k}}\left({W}_{\bf k}\hat{a}_{e{\bf k}}\hat{a}_{e{-\bf k}}+c.c\right)$. The ground state wave function can be written as $|G\rangle_p=\prod_{{\bf k}}e^{-\frac{\beta_{{\bf k}}}{\alpha_{{\bf k}}}\hat{a}^\dagger_{e{\bf k}}\hat{a}^\dagger_{e-{\bf k}}}|0\rangle$, where $ \alpha_{\bf k}=\frac{1}{2}\sqrt{\frac{\tilde{\epsilon}_{e{\bf k}}}{E_{\bf k}}+1}$, $\beta_{\bf k}=\frac{1}{2}\sqrt{\frac{\tilde{\epsilon}_{e{\bf k}}}{E_{\bf k}}-1}$, and $E_{e\bf k}=\sqrt{(\xi_{e{\bf k}}-\mu+4\psi_g^2\mathcal{U}^{ggee}_{00{\bf k}{\bf k}})^2-4|W_{\bf k}|^2\psi_g^4}$\cite{Pethick}.  There is a critical value $\mu_c$ that can be written as
\begin{equation}
\mu_c= \xi_{e{\bf k}_1}+(4\mathcal{U}^{eegg}_{00{\bf k}_1{\bf k}_1}-2|W_{{\bf k}_1}|)\psi_g^2,\label{muc}
\end{equation}
where ${\bf k}_1=\pi/d(\pm 1,0,0)$ is the location of the bottom of the excited band.  If $\mu<\mu_c$, the single-particle excitation in the excited band is gapped, i.e., $Min\{E_{e{\bf k}}\}>0$ as shown in Fig.(\ref{fig: fig2}A). As a result, $ \langle\hat{b}_{e{\bf R}}^\dagger \hat{b}_{e{\bf R}'}\rangle\sim e^{-|{\bf R}-{\bf R}'|/\chi}$ decays exponentially, where $\hat{b}_{e{\bf R}}=\int d{\bf k}\hat{a}_{e{\bf k}}e^{i{\bf k}\cdot{\bf R}}$ is the annihilation operator in real space and  the correlation length $\chi\sim (\mu-\mu_c)^{-1/2}$. Though there is no single-particle condensate in the excited band, it can be shown that $\langle\hat{b}_{e{\bf R}}\hat{b}_{e{\bf R}} \rangle=-\int d{\bf k}{\beta_{\bf k}}/{\alpha_{\bf k}}\sim \psi_g^2$, i.e., there is an off-diagonal long range order in the reduced two-particle density matrix for the excited band. Therefore we refer to it as a pair-condensate in the excited band. 

On the other hand, by substituting $\mu=\mu_c$ using Eq.(\ref{muc}) to the general expression for $E_{e\bf k}$ in the above paragraph, it is straightforward to show that $E_{e\bf k}=2|W_{{\bf k}_1}|\psi_g^2\sqrt{\xi_{e{\bf k}}-\xi_{e{\bf k}_1}}$ if $\mu=\mu_c$ is satisfied. When ${\bf k}\rightarrow {\bf k}_1$, $E_{e{\bf k}}\rightarrow 0$ and the dispersion becomes linear at the edge of BZ as shown in Fig.(\ref{fig: fig2}B).  The vanishing gap indicates that a condensate emerges with a finite momentum ${\bf k}_1$.  The condensate amplitude  $\psi_{e\pi}=\langle\hat{ a}_{e{\bf k}_1}\rangle$ can be calculated by minimizing the energy $ E=A|\psi_{e\pi}|^2+B|\psi_{e\pi}|^4$, where $A=(\xi_{e{\bf k}_1}-\mu+2\psi_g^2\mathcal{U}^{eegg}_{00{\bf k}_1{\bf k}_1})$, $B=\mathcal{U}^{eeee}_{{\bf k}_1{\bf k}_1{\bf k}_1{\bf k}_1}$, and $|\psi_{e\pi}|=-{A}/{2B}$. While excited-band condensates have also been found theoretically in a standard optical lattice before\cite{interband1,interband2}, a double-well lattice has the advantage that a very large scattering length of atoms is not required since the band gap is significantly reduced. For example, we found that for $V_L=5E_R$, $V_S=2E_R$, $V_z=15E_R$ and $a_s/d=0.04$, $\psi_{e0}^2/\psi_g^2$ is about $3\%$. 

As $\mu$ is determined by  $U_{g}$ and $\psi_g$ through the relation $\mu=\xi_{g0}+2U_g\psi^2_g$, it is worthwhile to discuss under what conditions  $\mu\ge\mu_c$ can be satisfied. For this purpose, we define $E_g=\xi_{e{\bf k}_1}-\xi_{g0}>0$. Note $W_{{\bf k}_1}=U^{eegg}_{00{\bf k}_1{\bf k}_1}$, $E_{e{\bf k}_1}$ can be rewritten as 
\begin{equation}
E_{e{\bf k}_1}=\sqrt{E_g^2+2E_g(4c-2)U_g\psi_g^2+(12c^2-16c+4)U_g^2\psi_g^4},
\end{equation}
where $c=U^{ggee}_{00{\bf k}_1{\bf k}_1}/U_g$. A straightforward analysis shows that if $c<1$, $E_{{\bf k}_1}$ vanishes at a sufficiently large value $\psi_g$. If $c\ge1$, the single-particle excitation gap for the excited band will not vanish, no matter how large $\psi_g$ is. The reason is that increasing $\psi_g$ has two competing effects on $E_{e{\bf k}}$. It first increases the chemical potential $\mu\sim U_g$ which favors reducing the gap. On the other hand, it also increase the intra-band density-density interaction $\mathcal{U}^{ggee}_{00{\bf k}-{\bf k}}$. In other words, it enhances the single-particle gap and increase the threshold for the chemical potential to close the gap. Which effect wins depends on the ratio between $U_g$ and $\mathcal{U}^{ggee}_{00{\bf k}{\bf k}}$. Though we  focus on the case where there is a large condensate in the lowest band in this work, our discussion can be easily generalized to the case where atoms in the lowest band form a Mott insulator instead of a condensate (see Supplementary Materials at [] for the discussions on Mott insulators). 


{\bf (C) Asymmetric lattices:} Because the inversion symmetry is broken in asymmetric lattices, $G_0$ becomes finite. It thus leads to an inter-band scattering process $\sim \hat{a}_{g0}^\dagger\hat{a}_{g0}^\dagger\hat{a}_{g0}\hat{a}_{e0}+c.c$.  We refer to this as an \textit{interaction induced inter-band hybridization}. In the mean-field description, it gives rise to the terms in Eq.(\ref{Ge}). 

The significance of $U'$ is that it leads to  a finite condense $\psi_{e0}=\langle \hat{a}_{e0}\rangle$ provided that $\psi_g\neq0$.  Unlike symmetric lattices where only the pairing term $\psi_g^2({W_{\bf k}}\hat{a}_{1{\bf k}}^\dagger\hat{a}_{1-{\bf k}}^\dagger+c.c)$ is present,  there is no threshold in the chemical potential for an asymmetric lattice to produce a condensate in the excited band. The only thing that matters is the ratio between $\tilde{\epsilon}_{e0}-\tilde{\epsilon}_{g0}$ and $G_0$, which  determines the amplitude of $\psi_{e0}$.  An asymmetric double-well lattice has two unique advantages. First, it reduces significantly the energy difference between the lowest two bands $\tilde{\epsilon}_{e0}-\tilde{\epsilon}_{g0}$,  similar to what was found in \cite{QZ}. Second, it provides  flexibility in tuning the value of $G_0$. As a result, it is possible to achieve a reasonably large condensate in the lowest excited band, as we show below. 

To study how the condensate forms at ${\bf k}=0$ in the excited band, we separate the following piece from the Hamiltonian in Eq.(\ref{He}), 
\begin{equation}
H_{e0}=\mathcal{K}_{e0}+\mathcal{V}_{e0}+(\psi_g^2W_0\hat{a}_{e0}^\dagger\hat{a}_{e0}^\dagger+\psi_g^3{G}_0\hat{a}_{e0}^\dagger+c.c)\label{He1}
\end{equation}
where $\mathcal{K}_{e0}=\tilde{\epsilon}_{e0}\hat{a}_{e0}^\dagger\hat{a}_{e0}$ and $\mathcal{V}_{e0}=\mathcal{U}_{0000}^{eeee}\hat{a}_{e0}^\dagger\hat{a}_{e0}^\dagger\hat{a}_{e0}\hat{a}_{e0}$. Without losing generality, a  repulsion $\mathcal{V}_{e0}$ has been added to the Hamiltonian.  We have also set both $u_{e0}$ and $u_{g0}$ to be real. There are two different driving terms for the excited band condensate. The quadratic term $\hat{a}_{e0}^\dagger\hat{a}_{e0}^\dagger$ favors the formation of a pair-condensate. The linear term $\hat{a}_{e0}^\dagger$, however, induces a single-particle condensate. The competition between them leads to an interesting evolution from a pair-condensate to a single-particle condensate in the excited band with increasing $G_0$. 

\begin{figure}[tbp]
\begin{center}
\includegraphics[width=3.2 in]{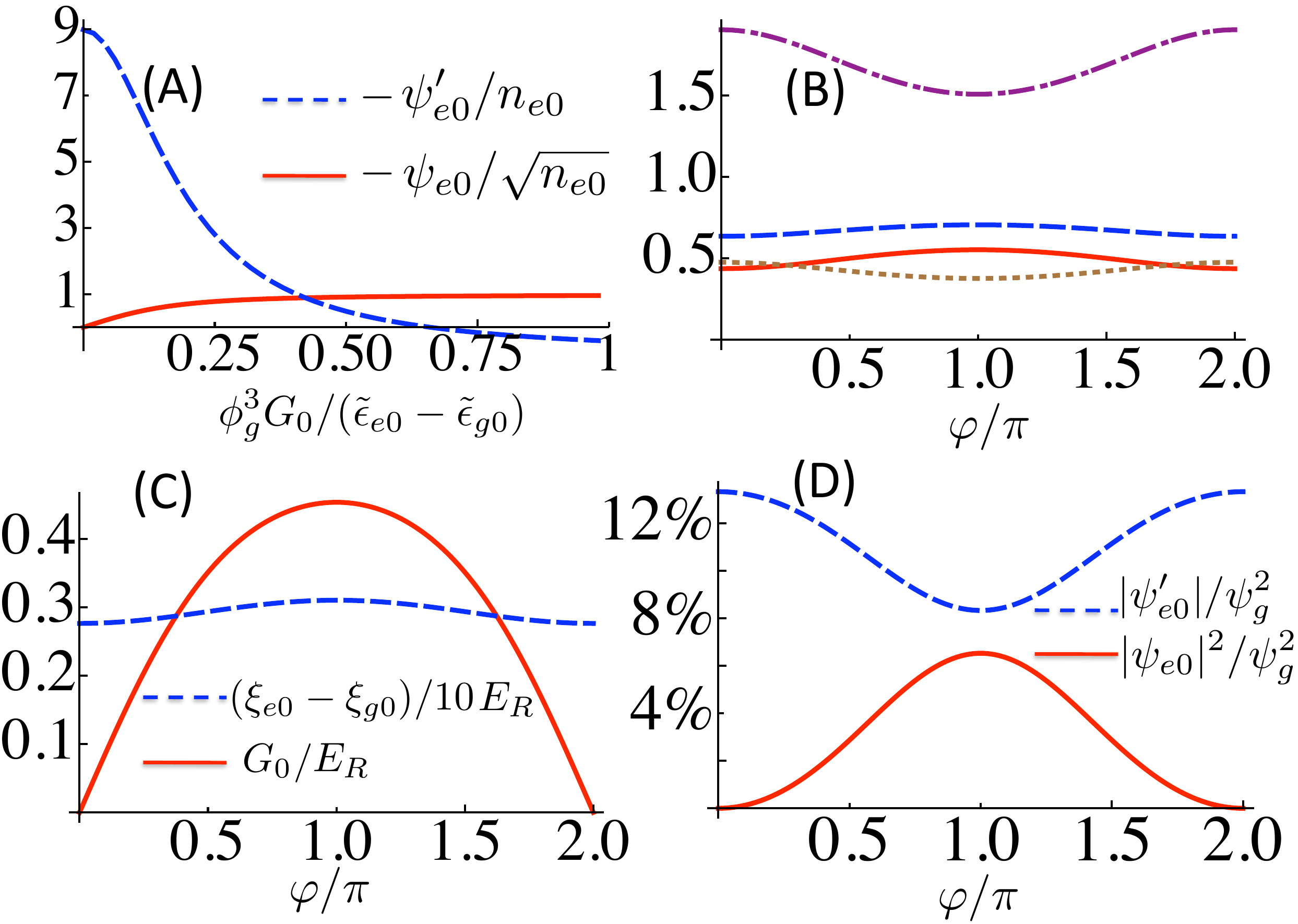}
\end{center}
\caption{(Color online)  (A) For an asymmetric double-well lattice, $\psi_{e0}$ and $\psi^{'}_{e0}$ as a function of $\psi_g^3G_0/(\tilde{\epsilon}_{e0}-\tilde{\epsilon}_{g0})$, where $W_0\psi_g^2/(\tilde{\epsilon}_{e0}-\tilde{\epsilon}_{g0})=0.2$ and $U_{e0}/(\tilde{\epsilon}_{e0}-\tilde{\epsilon}_{g0})=0.4$. (B) $U_{g}$(red solid), $U_{e0}$(blue dashed), $4\mathcal{U}_{0000}^{eegg}$(purple dash-dotted) and $W_0$(Brown dotted) as a function of $\varphi$ that is defined in Eq.(\ref{lp}) to characterize the relative position of the long and short lattice potential, where $V_L^y/E_R=V_L^z/E_R=15$, $V_L^x/E_R=2$, $V_S/E_R=6$, and $a_s/d=0.06$. (C) $G_0$ (red solid) and $\xi_{e0}-\xi_{g0}$(blue dashed)  as a function of $\varphi$. The parameters are the same as (B). (D) $\psi_{e0}^2/\psi_{g}^2$ for the same lattice configuration as (B) and (C).}\label{fig: fig3}
\end{figure}

The ground state for the Hamiltonian in Eq.(\ref{He1}) is written as $|G\rangle_s=\sum_nc_n\frac{\hat{a}^{\dagger n}}{\sqrt{n!}}|0\rangle$, where the coefficient $c_n$ can be calculated exactly. The order parameters for the single-particle and pair condensate  are defined by $\psi_{e0}=\langle \hat{a}_{e0}\rangle$ and $\psi'_{e0}=\langle \hat{a}_{e0}\hat{a}_{e0}\rangle$  respectively. An illuminating example to demonstrate how $\psi_{e0}$ and $\psi'_{e0}$ change as a function of $\psi_g^3G_0$ is shown in Fig.(\ref{fig: fig3}A). As seen from Eq.(\ref{He1}), the amplitude of $\psi_{e0}$ depends on the product of $\psi_g^2$ and $G_0$. On the other hand, it also depends on the value of  $\tilde{\epsilon}_{e0}-\tilde{\epsilon}_{g0}$.  The larger the latter is, the smaller the amplitude of $\psi_{e0}$ is.  Thus, we have chosen the dimensionless number $\psi_g^3G_0/( \tilde{\epsilon}_{e0}-\tilde{\epsilon}_{g0})$ as the $x$-axis in Fig.(\ref{fig: fig3}A).  When $G_0=0$, $\psi_{e0}=0$, we have a pure pair-condensate. The negative value of $\psi'_{e0}$ simply means that this pair condensate has a relative phase $\pi$ with respect to $\psi_g$, since $W_0>0$ in our case. With increasing $G_0$, $|\psi_{e0}|$ increase while $|\psi'_{e0}|$ decreases.  
In the large $G_0\psi_g^3$ limit, $\psi_{e0}/\sqrt{n_{e0}}\rightarrow 1$, where $n_{e0}=\langle \hat{a}_{e0}^\dagger\hat{a}_{e0}\rangle$. This indicates that the ground state for the Hamiltonian in Eq.(\ref{He}) becomes a coherent state. 

To concrete the above discussions for real cold atom experiments, we have carried out an exact band structure calculation for various double-well lattice configurations. The parameters in Eq.(\ref{He1}) for a typical lattice configuration are shown in Fig.(\ref{fig: fig3}B,C). The condensate amplitude in the excited band is  shown in Fig.(\ref{fig: fig3}D) for different values of $\varphi$. We have set $\psi_g^2=2$. It is clear that $\psi_{e0}^2$ can be as large as a few to ten percent of $\psi_g^2$. Thus for total particle number $N\sim 10^5-10^6$ in the weakly interacting region, the condensate particle number in the excited band can be of the order of $10^3-10^4$.  For the lattice configuration as shown in Fig(\ref{fig: fig3}B-D), the maximum of $\psi_{e0}^2$ occurs at $\varphi=\pi$, where $G_0$ also reaches its maximum as shown in Fig.(\ref{fig: fig3}C). We emphasize that this interaction induced condensate comes from the combination of interaction effect and the lattice potential that breaks the inversion symmetry. Without reducing the value of $\xi_{e0}-\xi_{g0}$, the condensate amplitude would be rather small though a finite coupling in principle can exist if the inversion parity is broken for any reasons. This interaction induced condensate in the excited band is a property of the ground state, since the interaction naturally mixed the condensates in the lowest two bands. Thus, it does not suffer from the short life time problem of the usual single excited band condensate.


{\bf (D) Probing condensates in different bands:} The crystal momentum distributions from different bands overlap with each other in the standard Time-Of-Flight(TOF) expansion. In a ballistic expansion, the interference pattern represents the original momentum distribution $\langle \hat{n}_{\bf k}\rangle=\int d{\bf R}_1d{\bf R}_2e^{i{\bf k}\cdot{({\bf R}_1-{\bf R}_2} )}\langle \hat{B}^\dagger_{{\bf R}_1}\hat{B}_{{\bf R}_2}\rangle $, where $\hat{n}_{\bf k}$ is the density operator in the momentum space and $\hat{B}_{{\bf R}}$ is the field operator in the real space. Expanding $\hat{B}_{\bf R}$ in the basis of Bloch wave functions, i.e., $\hat{B}_{\bf R}=\sum_{\sigma{\bf k}}\Psi_{\sigma{\bf k}}({\bf R})\hat{a}_{\sigma{\bf k}}$, one immediately sees that $\langle \hat{n}_{\bf k}\rangle=\sum_{\sigma\sigma'{\bf p}{\bf q}}\phi^*_{\sigma{\bf p}}({\bf k})\phi_{\sigma'{\bf q}}({\bf k})\langle \hat{a}^\dagger_{\sigma{\bf p}}\hat{a}_{\sigma' {\bf q}}\rangle$,  where $\phi_{\sigma{\bf q}}({\bf k})=\int d{\bf R}e^{-i{\bf k}\cdot{\bf R}}\Psi_{\sigma{\bf q}}({\bf R})\delta_{{\bf k}-{\bf q}={\bf K}}$ is the Fourier transform of the Bloch wave function. Therefore, the population at the reciprocal lattice vectors ${\bf k}_l=l({2\pi}/{d},0,0)$, where $l$ is an integer, can be written as 
\begin{equation}
n_{{\bf k}_l}=\sum_{\sigma=g,e}\phi^*_{\sigma 0}({\bf k}_l)\phi_{\sigma' 0}({\bf k}_l)\langle \hat{a}^\dagger_{\sigma 0}\hat{a}_{\sigma' 0}\rangle.
\end{equation}
Thus one can first numerically obtain the Bloch wave functions $\Psi_{\sigma{\bf k}}({\bf R})$ and its Fourier transforma $\phi_{\sigma{\bf q}}({\bf k})$ for the double-well lattice, and then use them to fit the TOF images. From the amplitude of a number of peaks in the interference pattern, one is able to extract $\langle \hat{a}^\dagger_{\sigma 0}\hat{a}_{\sigma' 0}\rangle$, i.e., the condensate amplitude at ${\bf k}=0$ of different bands. For the condensate at the edge of BZ, additional peaks located at ${\bf k}=\pi/d\hat{x}+l {\bf K}$ emerge as the unique feature of this condensate with a finite crystal momentum.  Another method is to apply the band mapping technique\cite{bm}, which can directly map out the crystal momentum distribution where condensates of different bands can be directly measured.

{\bf (E) Condensates in a superlattice:} Finally, we discuss the general principle of producing condensates in excited bands of a superlattice composed by multiple-well potentials in each single site. (1) A simple way to create a superlattice containing $n$ wells at each site is to use two lasers with wave lengths $\lambda$ and $n\lambda$. The lowest few bands may be well separated from other higher bands. (2) When this multiple-well potential respects inversion symmetry, the lowest band condensate automatically induces a condensate in an excited band of \textit{even parity }at ${\bf k}=0$, because of the presence of the interaction term $U'_1=G'\hat{a}^\dagger_{g0}\hat{a}^\dagger_{g0}\hat{a}_{g0}\hat{a}_{e'0}+c.c$ in the Hamiltonian, where $G'\neq 0$. For \textit{odd-parity} bands, $G'$ and $U'_1$ vanish. However, $U'_2=W'\hat{a}^\dagger_{g0}\hat{a}^\dagger_{g0}\hat{a}_{e'{-\bf k}_1}\hat{a}_{e'{\bf k}_1}+c.c$ is always present, and if the chemical potential is large enough to overcome the single-particle excitation gap for those bands,  condensates form at the edge of BZ with a finite momentum ${\bf k}_1$. When the inversion symmetry is broken for the asymmetric multiple-well potentials, $G'$ becomes finite for all the excited bands, and condensates in the lowest few excited bands can be induced. 

In summary, we have pointed out that in a double-well lattice, a stable condensate in the lowest excited band can be  induced by inter-band interactions. It either comes from large enough interaction that overcomes the single-particle excitation gap for the excited band, or a combination of inter-band interaction effect and the lattice potential that breaks the inversion symmetry. Our proposal can be generalized for producing condensates in the lowest few excited bands of a superlattice.  Recently, Sengstock's group has reported an observation of interaction-induced condensate in an excited band of a triangular double-well lattice\cite{Sengstock2}. This discovery directly proves the validity and  feasibility of our proposals. This work is supported by JQI-NSF-PFC, ARO-DARPA-OLE,  ARO-MURI and AFOSR-MURI.

\end{document}